\documentclass[a4paper,11pt]{article}
\usepackage[margin=2.5cm]{geometry}
\usepackage{setspace,graphicx,url,hyperref,enumitem}

\usepackage{amsmath,amssymb,amsfonts,amsthm,natbib}

\begin{document}
\onehalfspacing
\title{Solving large Minimum Vertex Cover problems on a quantum annealer}
\author{Elijah Pelofske\footnote{Los Alamos National Laboratory, Los Alamos, NM 87545, USA},
Georg Hahn\footnote{Lancaster University, Lancaster LA1 4YW, U.K.}, and Hristo Djidjev\footnotemark[1]}
\date{}
\maketitle

\begin{abstract}
We consider the minimum vertex cover problem having applications in e.g.\ biochemistry and network security. Quantum annealers can find the optimum solution of such NP-hard problems, given they can be embedded on the hardware. This is often infeasible due to limitations of the hardware connectivity structure. This paper presents a decomposition algorithm for the minimum vertex cover problem: The algorithm recursively divides an arbitrary problem until the generated subproblems can be embedded and solved on the annealer. To speed up the decomposition, we propose several pruning and reduction techniques. The performance of our algorithm is assessed in a simulation study.
\end{abstract}

\section{Introduction}
\label{sec:intro}
We consider the minimum vertex cover problem (abbreviated as MVC in the remainder of this article), one of Karp's 21 NP-complete problems. Formally, we are given an undirected graph $G=(V,E)$ with vertex set $V$ and edge set $E \subseteq V \times V$. A subset $V' \subseteq V$ is called a \textit{vertex cover} if every edge in $E$ has at least one endpoint in $V'$, that is, if for every $e=(u,v) \in E$ it holds true that $u \in V'$ or $v \in V'$. A \textit{minimum vertex cover} is a vertex cover of minimum size.

For NP-hard (graph) problems, such as MVC, graph partitioning or the maximum clique problem, novel technology allows us to obtain solutions which are very hard to find classically \cite{Chapuis2017}. One such device is the quantum annealer of D-Wave Systems, Inc.~\citep{TechnicalDescriptionDwave}, which allows us to find approximate solutions of quadratic unconstrained binary optimization and Ising problems, given by the minimization $\min H(x_1,\ldots,x_n)$, where
\begin{align}
    H(x_1,\ldots,x_n) = \sum_{i=1}^n a_i x_i + \sum_{i<j} a_{ij} x_i x_j.
    \label{eq:ising}
\end{align}
In \eqref{eq:ising}, the $a_i \in \mathbb{R}$, $i \in \{1,\ldots,n\}$, are linear weights and $a_{ij} \in \mathbb{R}$ for $i<j$ are quadratic couplers. The problem \eqref{eq:ising} is called a \textit{quadratic unconstrained binary optimization (QUBO)} problem if $x_i \in \{0,1\}$ and an \textit{Ising problem} if $x_i \in \{-1,+1\}$ for all $i$. The function \eqref{eq:ising} is often called a QUBO or Ising Hamiltonian, respectively. Both QUBO and Ising formulations are equivalent~\citep{Djidjev2016EfficientAnnealing}. The D-Wave quantum annealer aims to find a minimum of the function \eqref{eq:ising} by mapping it to a physical quantum system, from which a solution is read off after annealing is completed.

It is known that many NP-hard problems can be easily expressed as the minimization of a Hamiltonian of the form \eqref{eq:ising} including, for instance, graph partitioning, graph coloring, or maximum clique finding: see \cite{Lucas2014} for a comprehensive overview of QUBO and Ising Hamiltonians for a variety of NP-hard problems. According to \cite[Section~4.3]{Lucas2014}, the MVC Hamiltonian for a graph $G=(V,E)$ is given by
\begin{align}
    H = A \sum_{(u,v) \in E} (1-x_u)(1-x_v) + B \sum_{v \in V} x_v.
    \label{eq:MVC}
\end{align}
In \eqref{eq:MVC}, each $x_v \in \{0,1\}$ for $v \in V$ is a binary variable indicating if vertex $v$ belongs to the MVC. The first term in \eqref{eq:MVC} encodes the constraint that each edge in $G$ is incident to at least one vertex in the MVC, and the second term is simply the size of the MVC. As shown in \cite{Lucas2014}, choosing $0<B<A$ ensures that minimizing \eqref{eq:MVC} is equivalent to solving the MVC problem. As an explicit choice, we fix $B=1$ and $A=2$ in the remainder of the article.

Our goal in this paper is to compute the minimum of \eqref{eq:MVC} on a quantum annealer. However, this is often not possible without preprocessing due to a variety of reasons: First, there is a limitation on the input problem sizes we can solve due to the finite number of available qubits (up to roughly 5000 qubits for the newest D-Wave generation). Second, even if the number of qubits exceeds the number of variables, the current (D-Wave) technology provides only limited qubit connectivity \citep{TechnicalDescriptionDwave}. It is thus not guaranteed that all the required quadratic couplers in \eqref{eq:MVC} needed to map MVC onto the annealer hardware are available. This problem can be alleviated with a so-called minor embedding of the problem Hamiltonian onto the D-Wave hardware, where each variable is embedded onto a set of connected qubits, rather than a single one, at the expense, however, of a severe reduction in the number of available qubits~\citep{Chapuis2017}. For instance, the largest embeddable complete graph on D-Wave 2X has $46$ vertices, thus guaranteeing that arbitrary problems with $46$ variables can be approximated on D-Wave. 

In this article we propose a novel decomposition algorithm for MVC in order to allow larger problems to be solved on D-Wave (for instance, problem sizes exceeding the number of available qubits). The decomposition algorithm recursively splits a given instance of MVC into smaller subproblems until, at a certain recursion level, the generated subproblems are small enough to be solved directly, e.g., using a quantum annealer. The decomposition algorithm is exact, meaning that the optimality of the solution is guaranteed provided all generated subproblems are solved exactly. A variety of techniques outlined in Section~\ref{sec:pruning} allows us to eliminate entire subproblems during the recursion that cannot contribute to the MVC, or to reduce the size of subproblems by removing vertices which cannot belong to the MVC.

This article is structured as follows. After a brief literature review in Section~\ref{sec:literaturereview}, Section~\ref{sec:algorithm} introduces our decomposition algorithm for MVC. To prune subproblems during the decomposition, we discuss a variety of bounds on the size of the MVC in Section~\ref{sec:pruning}. We assess the performance of our decomposition method in a detailed simulation study in Section~\ref{sec:experiments}. The article concludes with a discussion of our results in Section~\ref{sec:discussion}.

\subsection{Literature review}
\label{sec:literaturereview}
The development of exact algorithms for NP-hard problems has been an area of constant attention in the literature \citep{dimacs1996, dimacs2000, Woeginger2008}.

In particular, the minimum vertex cover problem has been widely studied in the literature from a variety of aspects \citep{DowneyFellows1992, Balasubramanian1998, StegeFellows1999, Chen2000, Chen2001, NiedermeierRossmanith2003}. For instance, \cite{Chen2010} present a selection of techniques to reduce the size of an MVC instance and introduce a polynomial space and exponential time algorithm of the order of $O(1.27^k)$, where $k$ is the sought maximal size of the MVC, thus improving over the $O(1.29^k)$ algorithm of \cite{Niedermeier20070UB}. A variation of the MVC problem, the weighted MVC problem, is studied in \cite{Xu2016}.

Decomposition algorithms, such as the algorithm presented in this article, have already been suggested in \cite{Tarjan1985} and successfully applied to solve a variety of NP-hard problems (see \cite{Rao2008}) such as graph coloring.

A decomposition algorithm for another NP-hard graph problem, the maximum clique problem, has been proposed in \cite{Chapuis2017, Pelofske2019}. In \cite{Pelofske2019}, the authors additionally investigate a variety of techniques to prune subproblems during the recursive decomposition, for instance by computing bounds on the clique size. Similarly, to solve the maximum independent set problem, an equivalent formulation of the maximum clique problem, several algorithms are known including some relying on graph decomposition \citep{Giakoumakis1997,Courcelle2000}.

\section{A decomposition method for MVC}
\label{sec:algorithm}
\begin{figure}[t]
    \centering
    \includegraphics[width=0.5\textwidth]{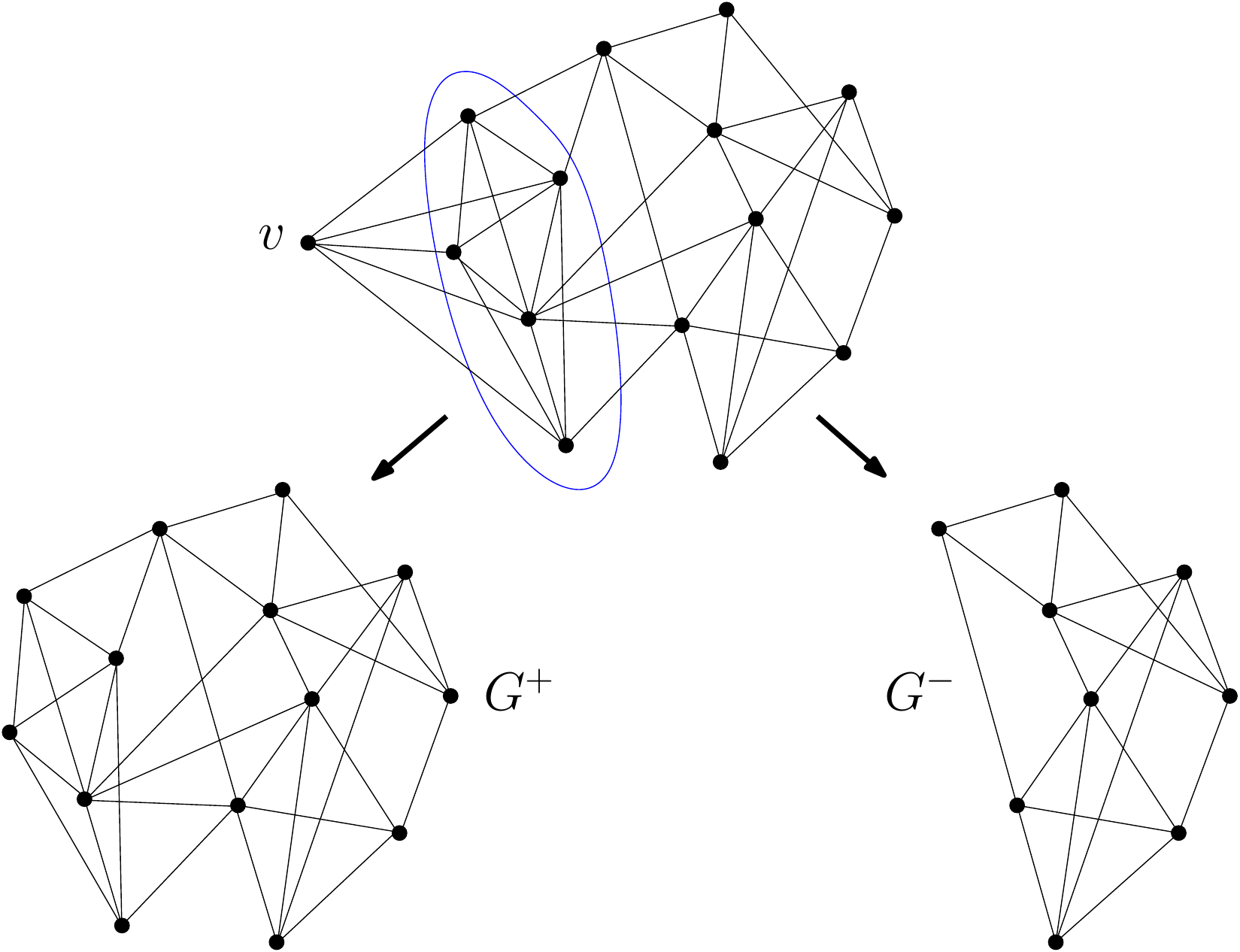}
    \caption{Illustration of the vertex splitting at a vertex $v$.}
    \label{fig:vertex_splitting}
\end{figure}
The focus of this article lies on a decomposition algorithm for MVC instances presented in this section.

We are given a graph $G=(V,E)$ for which we aim to compute a MVC. Let $V' \subseteq V$ denote the (unknown) MVC. If $G$ cannot be processed as a whole, we suggest the following technique to split the problem into two subproblems. Select any vertex $v \in V$. Then there are two cases: Either $v \in V'$ or $v \notin V'$, each case leading to a subproblem of reduced size (see Figure~\ref{fig:vertex_splitting}, top).

If $v \in V'$, we augment the set of vertices belonging to the MVC by $v$; afterwards, we can remove $v$ and all edges adjacent to $v$ since those edges are already covered by the choice of $v$. This is illustrated in Figure~\ref{fig:vertex_splitting} with subgraph $G^+$.

If $v \notin V'$, we observe that for all edges with endpoint $v$, that is all $(v,u) \in E$, it must hold true that $u \in V'$. This is true since, if $v \notin V'$, those edges must still be covered by their other endpoint $u$ in the MVC. Also, we can remove $v$ from $G$ since we know it is not in the MVC (likewise we can remove all $u$ with $(u,v) \in V$ and the adjacent edges of all such $u$ since those vertices are known to belong to the MVC). In Figure~\ref{fig:vertex_splitting}, under the assumption that $v \not\in V'$, all vertices inside the blue circle must belong to the MVC. After removing $v$ and all its adjacent edges, and assigning all encircled vertices to the MVC, we are left with the subgraph $G^-$.

Since the cases $v \in V'$ and $v \notin V'$ are exhaustive, we split $G$ into both $G^+$ and $G^-$ and continue to compute MVC on both subgraphs. If either graph is still too large to compute an (exact) solution of MVC on it, the above decomposition can be recursively applied. In a recursive application, some bookkeeping is needed to remember the current set of cover vertices for each generated subgraph.

Since neither of the generated subgraphs $G^+$ and $G^-$ contains $v$, the graph size is reduced by at least one in each recursion level, thus guaranteeing the termination of the algorithm.

The algorithm as described will output the exact MVC given we employ an exact method to solve MVC on the subgraphs at leaf level. Typically, for instance when applying D-Wave at leaf level to compute MVC solutions for graphs of at most $46$ vertices (see Section~\ref{sec:intro}), an exact solution is not guaranteed. However, our algorithm can also be applied probabilistically: If the solver applied to the subgraphs at leaf level finds optimal solutions with probability $p$, our decomposition algorithm will report the correct MVC for the original graph with the same probability $p$.

\section{Pruning techniques for MVC}
\label{sec:pruning}
The recursive decomposition proposed in Section~\ref{sec:algorithm} can be simplified and accelerated using a variety of bounding and pruning techniques.

\subsection{Upper and lower bounds}
\label{sec:bounds}
First, we bound the size of the MVC of each generated subgraph. If a lower bound on any subgraph exceeds the current best vertex cover size, we do not need to consider that subproblem for further decomposition as it cannot contain an optimal solution. We compute the following \textit{lower bounds}:
\begin{enumerate}[label=(\roman*)]
    \item First, we take the maximum of three easy to compute deterministic lower bounds.
        \begin{itemize}
        \item The function \textit{min\_weighted\_vertex\_cover} of the NetworkX package \citep{Networkx} computes an approximate vertex cover of at most twice the size of the optimal cover using the algorithm of \cite{BarEve85}. Hence, dividing its result by a factor of two results in a lower bound on the size of the MVC. 
        \item We employ the matrix rank upper bound on the order of the maximum independent set of \cite{Budinich2003} to any generated subgraph. Since  the complement of any independent set of vertices is a vertex cover, any  upper bound on the maximum independent set size corresponds to a lower bound on the MVC size.
        \item We use the easy to compute \textit{minimum degree bound} of \cite[page~20]{Willis2011}.
        \end{itemize}
    \item Any graph coloring provides an upper bound on the chromatic number. Since all vertices in a maximum clique need to be assigned different colors, the chromatic number is again an upper bound of the clique number. This means that it gives an upper bound of the size of the maximum independent set of the complement graph and can thus be used as described in the previous point. To compute a graph coloring we use the heuristic function \textit{greedy\_color} of the NetworkX package \citep{Networkx}, which is applied to the complement $\overline{G}$ of $G$. 
    \item Lastly, we compute the Lov\'asz number \citep{Lovasz1979} of $G$ if the size of $G$ is less than or equal to $50$. If $G$ is larger than $50$ we return instead the size of $G$, a trivial upper bound of the Lov\'asz number. The reason for this is the excessive computation time associated with finding the Lov\'asz number, as well as practical program memory limitations with the implementation we adapted from \cite{Stahlke2013}. The Lov\'asz number is an upper bound for the maximum independent set number of $G$ \citep{GroetschelLovaszSchrijver1988}, and consequently a lower bound on the size of the MVC of $G$. 
\end{enumerate}

We compute \textit{upper bounds} on the size of the MVC as follows:
\begin{enumerate}[label=(\roman*)]
    \item At any point in the decomposition tree, the best solution found so far in $G^+$ (the larger subgraph of the two generated subgraphs $G^+$ and $G^-$) is an upper bound on the size of the MVC. Therefore, we follow the recursive $G^+$ branches in the decomposition first until we can compute the MVC on any of the generated $G^+$ subgraphs: its MVC size then becomes a good upper bound for all the other (smaller) generated subgraphs. We call this strategy the \textit{decomposition bound}.
    \item The size of the MVC of $G$ added to the clique number of $\overline{G}$ equals the size of $G$. Consequently, any heuristic solver that finds an approximate solution of the clique number in $\overline{G}$ can be used to compute an upper bound on the size of the MVC of $G$. We apply the \textit{fmc solver} \citep{fmc} in heuristic mode to $\overline{G}$, thus giving us an upper bound on the size of the MVC of $G$.
\end{enumerate}

The above bounds are employed during the recursion to prune those subproblems which cannot contain vertices belonging to the MVC of the input graph.

\subsection{Reduction techniques}
\label{sec:reductions}
In addition to using upper and lower bounds, we also use three \textit{reduction techniques} that allow us to figure out a partial solution to the MVC by deciding for a subset of the vertices of $G$ whether they belong to $V'$ or not. These reduction techniques are the following:
\begin{enumerate}[label=(\roman*),wide]
    \item We coin the first method \textit{neighbor based vertex removal}. Essentially, we search and remove triangles, vertices of degree 1, and vertices of degree zero in any subgraph. Since in a triangle, any two arbitrarily chosen degree 2 triangle vertices belong to the MVC, we add a contribution of 2 to the overall size of the MVC and remove the triangle. Analogously, vertices of degree 1 are automatically in the MVC and can be removed, along with the only neighbor of that vertex, after adding a contribution of 1 to the current size of the MVC. Vertices of degree zero can be removed without further processing.
    \item Another reduction technique works on the QUBO formulation of the MVC given in \eqref{eq:MVC} instead of on the graph itself. In particular, for any subgraph produced by our algorithm we generate the corresponding MVC Hamiltonian \eqref{eq:MVC}, which is then analyzed. Several general-purpose preprocessing techniques are capable to identify binary relations or \textit{persistencies} in QUBO or Ising Hamiltonians. Persistencies determine the value of certain variables in every global minimum (strong persistencies) or at least one global minimum (weak persistencies). In \cite{Boros2002}, a comprehensive overview of such techniques is given. Suppose variable $x_v$ for vertex $v$ (see Section~\ref{sec:intro}) is assigned the value $x_v=1$ in the persistency analysis: we can then add $v$ to the current vertex cover and remove $v$ and its adjacent edges from the subgraph. If $x_v=0$ we can remove $v$ and its adjacent edges without further processing. We employ the \textit{qpbo} Python bindings of \cite{Rother2007} to carry out the persistency analysis.
    \item In \cite{AkibaIwata2015}, the authors state a variety of reduction techniques for MVC that have been used in the theoretical study of exponential-complexity branch-and-reduce algorithms. For instance, those techniques include \textit{degree-one reductions}, \textit{decomposition}, \textit{dominance rules}, \textit{unconfined vertex reduction}, \textit{LP-} and \textit{packing reductions}, as well as \textit{folding-}, \textit{twin-}, \textit{funnel-} and \textit{desk reductions}. We employ only the reduction methods from the Java package \textit{vertex\_cover-master} \citep{javaVC}. For a given input graph, \textit{vertex\_cover-master} returns a superset of the MVC. This means that any vertex not contained in the output of \textit{vertex\_cover-master} is definitely not part of the MVC and can be removed. The code can be applied repeatedly until no further vertices are found that can be removed.
\end{enumerate}

The simulations in Section~\ref{sec:experiments} assess the effectiveness of the aforementioned techniques.

\section{Experimental results}
\label{sec:experiments}
This section presents three performance analysis experiments. In Section~\ref{sec:evaluation}, we evaluate four different choices for the selection of the vertex $v$ used in the algorithm of Section~\ref{sec:algorithm} to split any subgraph into two smaller parts. Moreover, in Section~\ref{sec:evaluation}, we separately evaluate the bounds on the size of the MVC discussed in Section~\ref{sec:bounds}, and the reduction techniques of Section~\ref{sec:reductions}.

Based on the assessment in Section~\ref{sec:evaluation}, we select the best strategy of vertex selection for splitting, lower and upper bounds on the MVC size of the generated subgraphs, and reduction techniques, and call the resulting method the \textit{DBR} algorithm (decomposition, bounds, reduction). Section~\ref{sec:DBR} evaluates our DBR algorithm on simulated graphs, benchmark test graphs, and real world graphs. In Section~\ref{sec:future}, we make a prediction on its behaviour on future D-Wave architectures.

We compare all algorithms using two measures, the preprocessing time, as well as the solution time. The preprocessing time is the CPU clock time our algorithm needs to decompose a problem instance into subproblems of size $46$; the time to solve the subproblems then depends on the solver being used, and is hence not included in the preprocessing time. We define the solution time as the sum of the preprocessing time and an additional contribution of $1.6$sec*$N$, where $1.6$ seconds is the average QPU access time for $10^4$ anneals on D-Wave 2X, and $N$ is the number of leaf subgraphs generated in a decomposition run. 

All figures have logarithmic scale on the $y$-axis.

\subsection{Evaluation of vertex selection, bounding and pruning techniques for MVC}
\label{sec:evaluation}
\begin{figure}[t]
    \centering
    \includegraphics[width=0.49\textwidth]{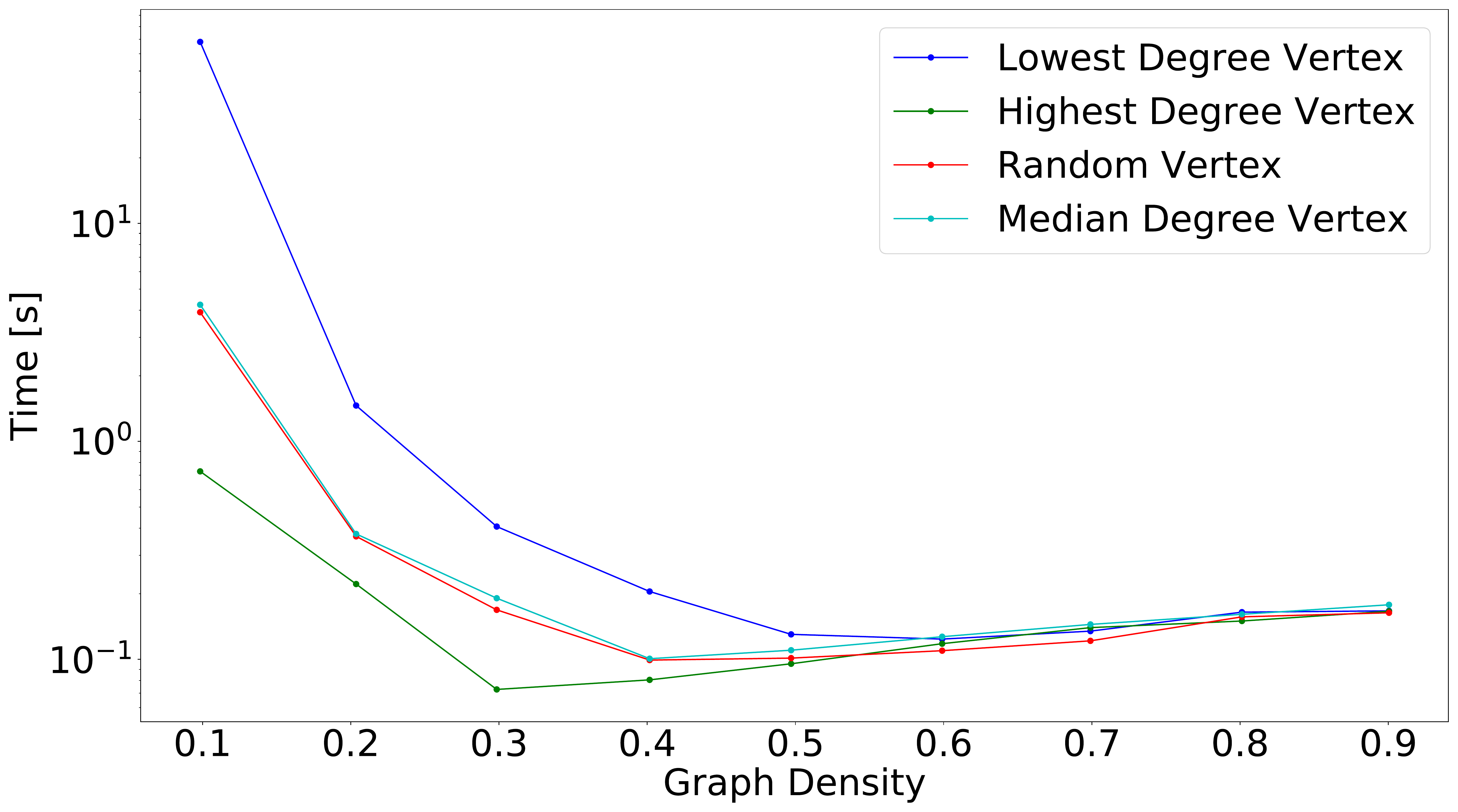}\hfill
    \includegraphics[width=0.49\textwidth]{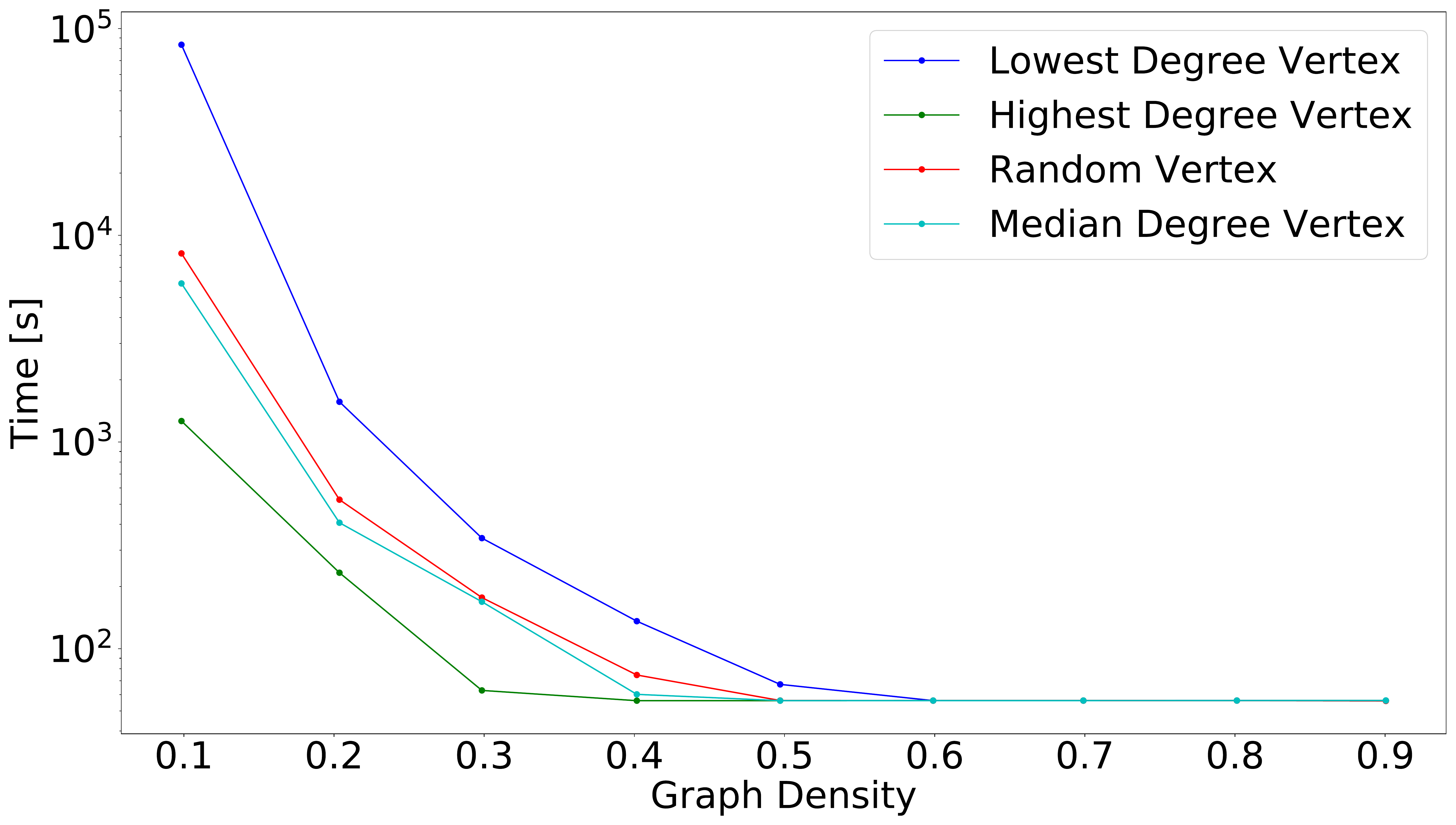}
    \caption{Preprocessing time (left) and solution time (right) for different vertex selection strategies. \label{fig:vertex_choice}}
\end{figure}
First, we compare four different options for the selection of the vertex $v$ used in Section~\ref{sec:algorithm} to split any subgraph into two parts during the decomposition. Those choices are the selection of a vertex of lowest degree, of highest degree, of median degree, and a random vertex. If several vertices exist in each category, $v$ is chosen at random among them.

For all figures presented in this section, we employ graphs of size $|V|=80$ having a graph density $d$ ranging from $0.1$ to $0.9$ in steps of $0.1$.

Experimental results for the vertex choice are presented in Figure~\ref{fig:vertex_choice} and show preprocessing time (left plot) and solution time (right plot). Selecting the highest degree vertex yields the both the lowest preprocessing and solution time. We therefore use the highest degree vertex selection in our implementation of the algorithm of Section~\ref{sec:algorithm}.

\begin{figure}[t]
    \centering
    \includegraphics[width=0.49\textwidth]{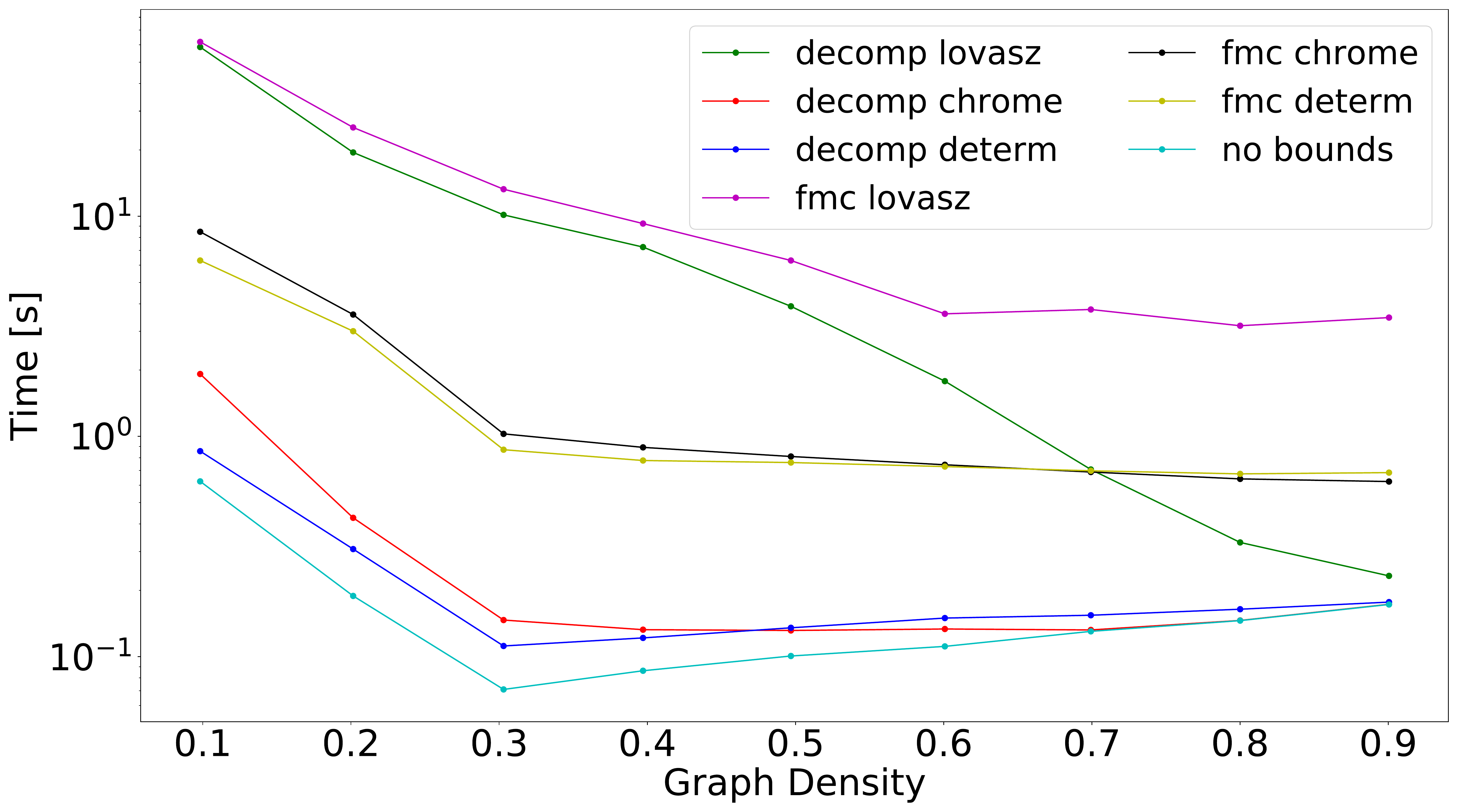}\hfill
    \includegraphics[width=0.49\textwidth]{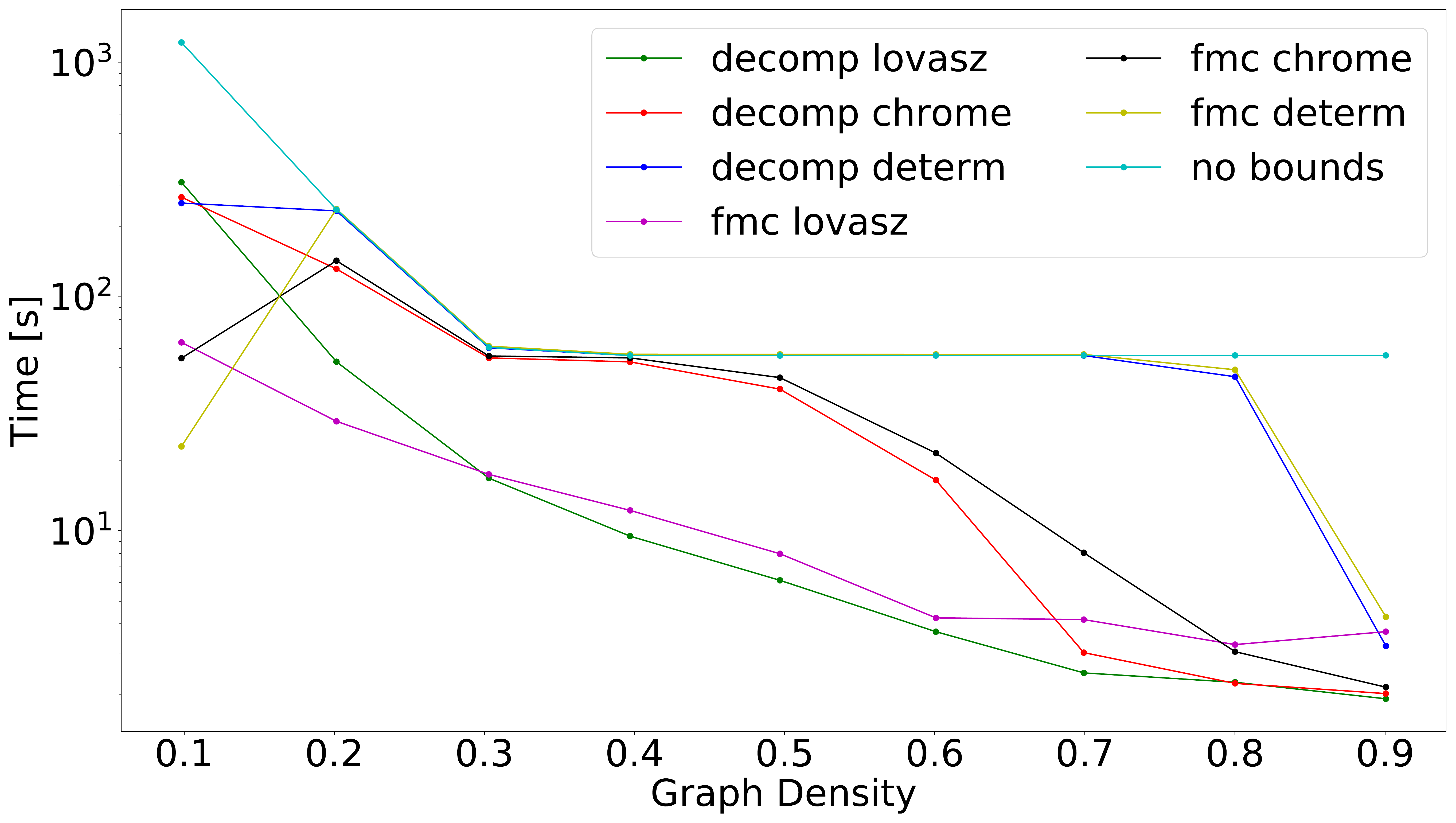}
    \caption{Preprocessing time (left) and solution time (right) for different combinations of lower and upper bounds. \label{fig:bounds}}
\end{figure}

Next, we test all six different options for combining the three lower and the two upper bounds presented in Section~\ref{sec:bounds} (against no bounds as seventh option). Here the picture is more complex (Figure~\ref{fig:bounds}). With respect to the preprocessing time, the version with "no bounds" is the best, and close to it are the combinations labeled "decomposition--deterministic" (the decomposition upper bound and the lower deterministic bound) and "decomposition-chrome" (the decomposition upper bound and the chromatic number lower bound). The two Lov\'asz-based combinations take the longest preprocessing times. Although they produce the lowest number of subgraphs, they also have the highest computational cost per bound (due to the computationally intensive computation of the Lov\'asz number already pointed out in \cite{Pelofske2019}). In Figure~\ref{fig:bounds} (left) we include in the total time only the preprocessing (CPU) time needed to decompose the original graph into subgraphs of sufficiently small sizes, but do not include the D-Wave time for solving these subproblems. In the next experiment, we do include that time. In Figure~\ref{fig:bounds} (right), we present solution time. 

There is no clear winner in Figure~\ref{fig:bounds} (right). The combinations "fmc-determ" and "fmc-chrome" perform best for the most difficult case, at 0.1 density. From density 0.2 the two Lov\'asz-based algorithms perform the best. The combination "fmc-deterministic" does not work well. At density 0.8 and 0.9, the combination "fmc-chrome" behaves similarly to the Lov\'asz-based options. Taking into account both experiments, we concluded that "fmc-chrome" offers the most balanced performance and we use this combination in our implementation of the algorithm of Section~\ref{sec:algorithm}. Although the bounds based on the Lov\'asz number give excellent subgraph reduction, both the calculation of the Lov\'asz number as well as our implementation becomes intractable to use in practice above the size limit ($50$) we used in Figure~\ref{fig:bounds}. Scaling this approach to larger D-Wave architectures as well as larger problem graphs would be entirely impractical for any approach involving the computation of the Lov\'asz number. Therefore, the best bound combination in terms of scalability as well as overall solution time is the decomposition upper bound in connection with the chromatic number lower bound. We will use this combination in the DBR algorithm in Section~\ref{sec:DBR}.

\begin{figure}[t]
    \centering
    \includegraphics[width=0.49\textwidth]{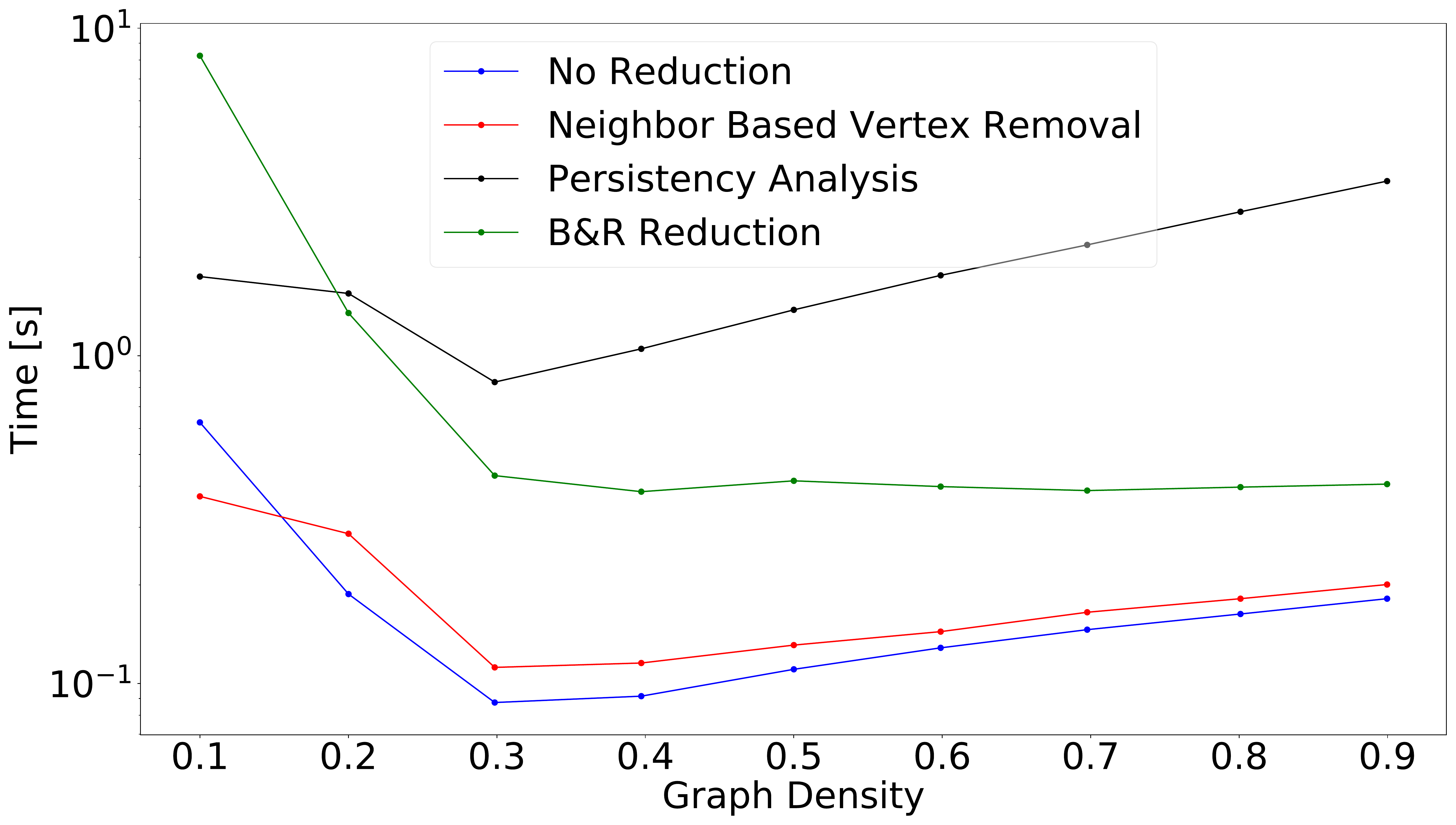}\hfill
    \includegraphics[width=0.49\textwidth]{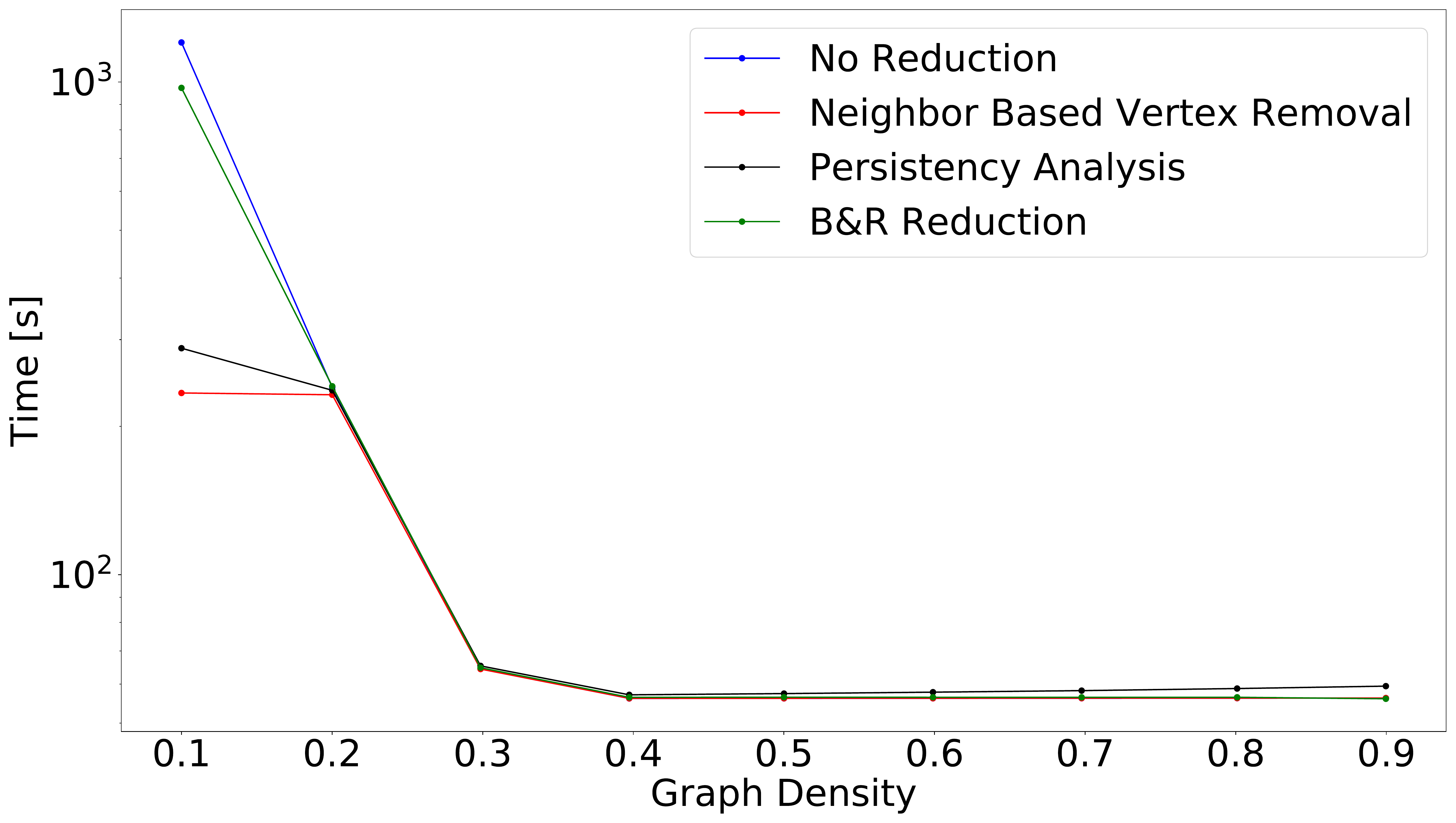}
    \caption{Preprocessing time (left) and solution time (right) for different reduction techniques.  \label{fig:reduction}}
\end{figure}
Third, we evaluate the reduction techniques of Section~\ref{sec:reductions}, where we apply any given reduction technique on every subgraph generated at each level of the decomposition tree. Results are shown in Figure~\ref{fig:reduction}. We observe that all three reduction techniques are capable of decreasing the number of generated subgraphs for low densities (in comparison to using no reduction in the decomposition algorithm), with the neighbor based vertex removal yielding the best results. Moreover, the neighbor based vertex removal is also the fastest technique, thus making it our choice for the DBR algorithm presented in Section~\ref{sec:DBR}.

In all experiments, one can observe that upon increasing the graph density (and thereby the size of the edge set of $G$) the running time decreases, which seems counter-intuitive. This can be explained by the specifics of our decomposition algorithm, which benefits from vertices of high degrees. In Figure~\ref{fig:vertex_splitting}, the larger the degree of the vertex $v$, the smaller the size of graph $G^-$. This results in decomposition trees with some very short branches for dense graphs, which in some cases may lead to nearly linear recursions. In contrast, for sparse graphs the decomposition trees are balanced and thus might be of exponential size.

\subsection{The DBR algorithm}
\label{sec:DBR}
We use our results from Section~\ref{sec:evaluation} in order to fully specify our decomposition algorithm presented in Section~\ref{sec:algorithm}. In particular, we employ the decomposition from Section~\ref{sec:algorithm} with highest degree vertex selection at each level of the decomposition. For each generated subproblem, we bound the size of the MVC it contains using the decomposition upper bound and the chromatic number lower bound (see Section~\ref{sec:bounds}).

\begin{figure}[t]
    \centering
    \includegraphics[width=0.49\textwidth]{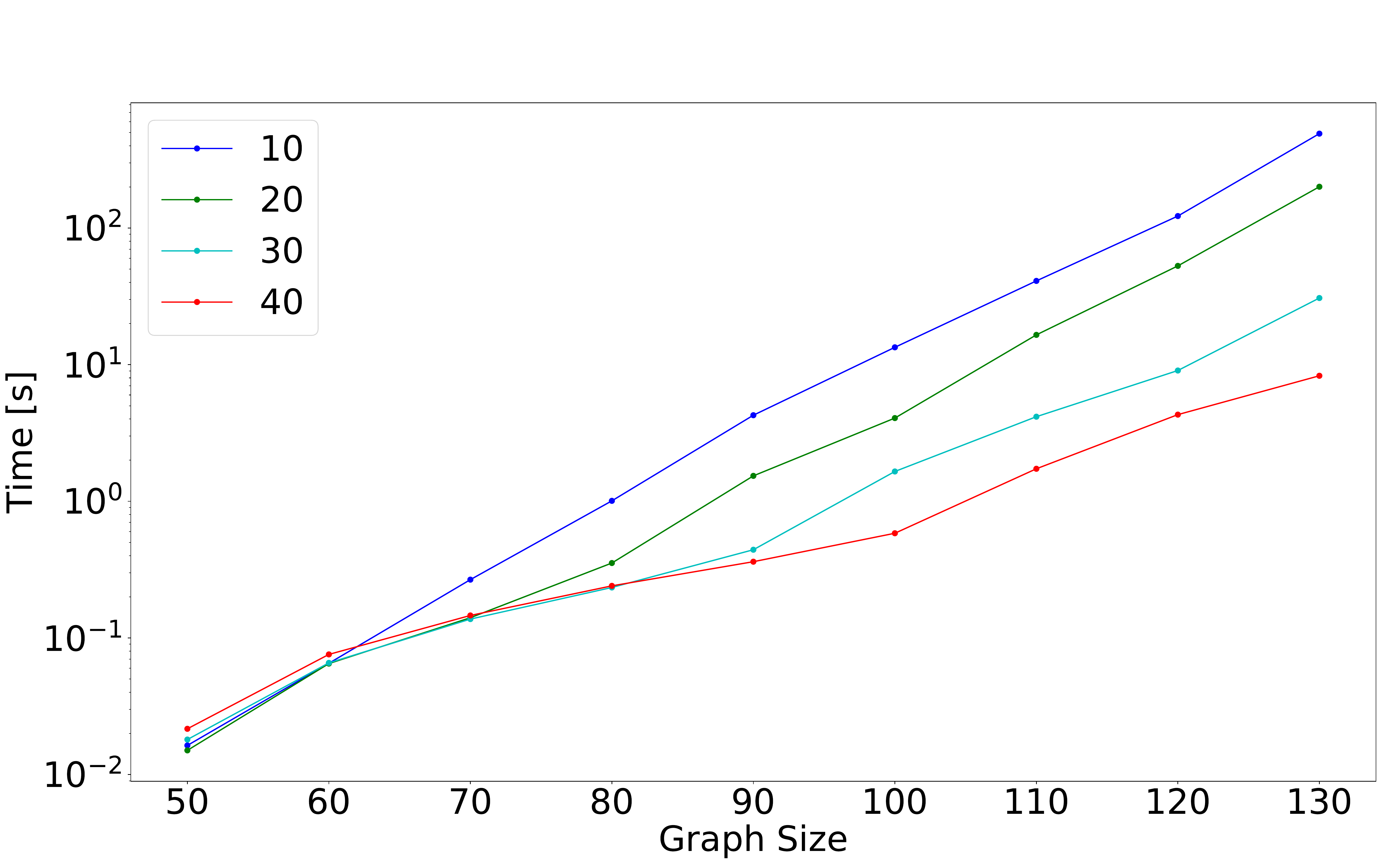}\hfill
    \includegraphics[width=0.49\textwidth]{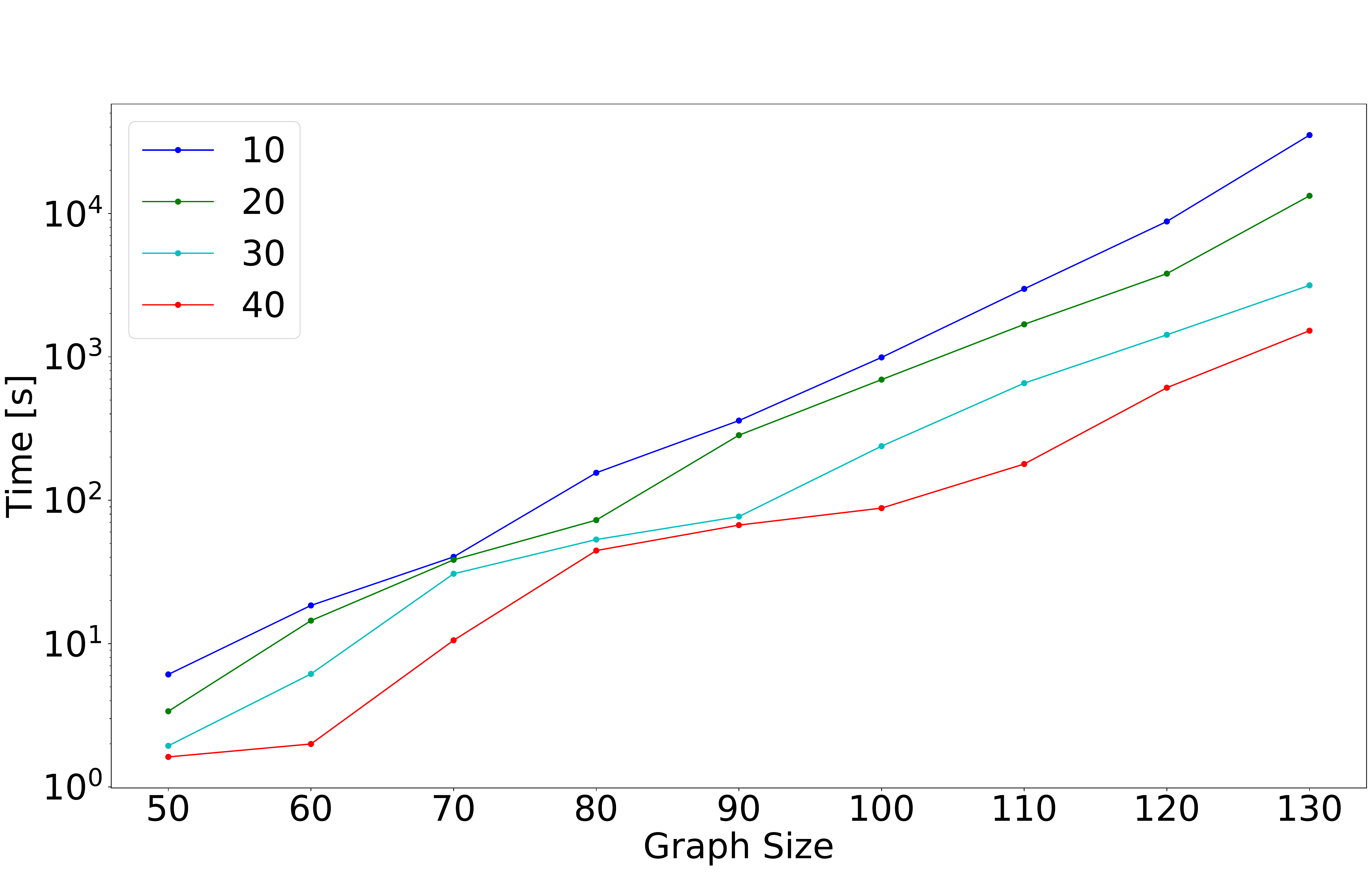}
    \caption{DBR algorithm applied to random graphs with average vertex degree $d \in \{10, 20, 30, 40\}$. Preprocessing time (left) and solution time (right). \label{fig:randomgraphs}}
\end{figure}
We apply the DBR algorithm to random graphs of size $|V| \in \{ 50,60,\ldots, 130\}$ and record the preprocessing time as well as the solution time. Figure~\ref{fig:randomgraphs} shows results for graphs with four fixed average vertex degrees $d \in \{10, 20, 30, 40\}$. Those results are based on the average of $10$ repetitions.

Due to the logarithmic scale on the $y$-axis, the results suggest an exponential increase in both subgraph count and computation time. This is to be expected as the MVC is an NP-hard problem. As outlined above, graphs with higher average vertex degrees yield lower numbers of generated subgraphs and thus faster computation times due to the characteristics of our decomposition algorithm.

\begin{table}[t]
\centering
\begin{tabular}{|l||l|l||l|l|l|}
    \hline
    Graph name ~ & No.~vertices~ & No.~edges~ & CPU time~ & No.~subgraphs~ & solution time~ \\
    \hline
    keller5 & 776 & 225990 & 672.54 & 654 & 1718.94 \\
    keller4 & 171 & 9435 & 2.24 & 30 & 50.24 \\
    mann-a45 & 1035 & 533115 & 842.42 & 1 & 844.02 \\
    p-hat300-2 & 300 & 21928 & 172.85 & 4276 & 7014.45 \\
    p-hat300-3 & 300 & 33390 & 12.35 & 195 & 324.35 \\
    p-hat500-3 & 500 & 93800 & 115.56 & 425 & 795.56 \\
    p-hat700-3 & 700 & 183010 & 524.76 & 1380 & 2732.76 \\
    brock200-1 & 200 & 21928 & 4 & 104 & 170.4 \\
    brock200-2 & 200 & 9876 & 31.51 & 715 & 1175.51 \\
    brock200-3 & 200 & 12048 & 8.26 & 134 & 222.66 \\
    gen400-p0.9-55 & 400 & 71820 & 33.62 & 3 & 38.42 \\
    c-fat200-5 & 200 & 8473 & 41.78 & 131 & 251.38 \\
    hamming8-4 & 256 & 20864 & 20.28 & 217 & 367.48 \\
    \hline 
\end{tabular}
\bigskip
\caption{Total solution time in seconds for DIMACS benchmark graphs based on a single run.\label{table:benchmark}}
\end{table}

Apart from random graphs, we also evaluate the DBR algorithm on DIMACS benchmarks graphs \citep{dimacs}. Results are given in Table~\ref{table:benchmark}. We observe that the DBR algorithm is capable of computing the optimal solution of MVC on such benchmark graphs in reasonable time.

We also evaluate the DBR algorithm on several real world graphs provided in the Network repository \citep{nr} in a single run. We specifically selected graphs with undirected and unweighted edges, and of a size of at most a few thousand vertices.

We selected graphs from the following categories in the Network repository (listed in the order they appear in Table~\ref{table:real_graphs}):
\begin{enumerate}
    \item the first five graphs in Table~\ref{table:real_graphs} belong to the \textit{Brain networks} category \citep{bigbrain},
    \item the next two graphs contain data from the \textit{Retweet networks} category \citep{rossi2014pmc-www, rossi2012fastclique},
    \item followed by two enzyme structure graphs in the \textit{Cheminformatics} category,
    \item two graphs from the \textit{Web graphs} category \citep{adamic2005political, gleich2004fast},
    \item two graphs from the \textit{Collaboration networks} category \citep{newman2006finding, de2011exploratory},
    \item two graphs from the \textit{Interaction networks} category \citep{cohen2005enron, opsahl2009clustering},
    \item one graph from the \textit{Infrastructure networks} category \citep{opsahl2011anchorage},
    \item one graph from the \textit{Biological networks} category \citep{duch2005community},
    \item one graph each from the category \textit{E-mail networks} and \textit{Proximity networks},
    \item one graph from the \textit{Technological networks} category \citep{rocketfuel},
    \item and a graph from the \textit{Road networks} category.
\end{enumerate}

Apart from applying the DBR algorithm, we solve each graph with the \textit{fmc solver} of \cite{fmc} (since fmc finds maximum cliques, we determine the vertex cover as those vertices not belonging to the maximum clique of the complement of $G$). Another way of solving for the optimal vertex cover is to apply a linear programming solver such as Gurobi \citep{gurobi} or CPLEX \citep{cplex} to the Hamiltonian in eq.~\eqref{eq:MVC}. However, as shown in \cite[Table~1]{Chapuis2019} for an equivalent classical NP-complete problem, using the fmc solver (which is tailored to graph problems) is considerably faster than employing a general-purpose solver such as Gurobi, and we thus report results for fmc only.

Results are given in Table~\ref{table:real_graphs}. We never observed in our simulations that the fmc solver was able to compute the optimal vertex cover when our DBR algorithm could not. We therefore only display graphs which could be solved by the DBR algorithm in reasonable time (sixth column of Table~\ref{table:real_graphs}), and indicate the solution time (or NA if fmc had not produced a result after 19 hours of wall clock runtime) for fmc in the last column. Table~\ref{table:real_graphs} suggests two main conclusions: First, the fmc solver is usually only able to compute the vertex cover for graphs with relatively few vertices. If fmc is able to compute a solution, it does so around 100 times faster than the DBR algorithm. Second, for graphs unsolvable with fmc, our DBR algorithm is able to compute the vertex cover in seconds or minutes, with a maximal solve time of around 15 minutes which seems reasonable for practical applications.

\begin{table}[t]
\centering
\scriptsize
\begin{tabular}{|l||l|l||l|l|l||l|}
    \hline
    Graph name ~ & No.~vertices~ & No.~edges~ & CPU time~ & No.~subgraphs~ & solution time~ & fmc solver~ \\
    \hline
    bn-macaque-rhesus-interareal- & 93 & 2700 & 0.25 & 1 & 1.85 & 0.018 \\
    \hfill cortical-network-2 &&&&&&\\
    bn-mouse-visual-cortex-2 & 193 & 214 & 0.01 & 1 & 1.61 & 0.104 \\
    bn-macaque-rhesus-cerebral- & 91 & 1401 & 0.14 & 1 & 1.74 & 0.022 \\
    \hfill cortex-1 &&&&&&\\
    bn-macaque-rhesus-brain-2 & 91 & 582 & 0.01 & 1 & 1.61 & 0.023\\
    bn-macaque-rhesus-brain-1 & 242 & 3054 & 23.31 & 427 & 706.51 & NA \\
    rt-retweet & 96 & 117 & 0.05 & 1 & 1.65 & 0.030 \\
    rt-twitter-copen & 761 & 1029 & 0.94 & 1 & 2.54 & NA \\
    ENZYMES-g296 & 125 & 141 & 0.17 & 1 & 1.77 & NA \\
    ENZYMES-g103 & 59 & 115 & 0.02 & 1 & 1.62 & 0.012\\
    web-polblogs & 643 & 2280 & 6.4 & 1 & 8 & NA \\
    web-edu & 3031 & 6474 & 51.35 & 518 & 880.15 & NA \\
    ca-netscience & 379 & 941 & 1.71 & 66 & 107.31 & NA \\
    ca-CSphd & 1882 & 1740 & 6.53 & 1 & 8.13 & NA \\
    ia-enron-only & 143 & 623 & 0.99 & 10 & 16.99 & NA \\
    ia-fb-messages & 1266 & 6451 & 3.71 & 1 & 5.31 & NA \\
    inf-openflights & 2939 & 15677 & 7.87 & 1 & 9.47 & NA \\
    bio-celegans & 453 & 2025 & 0.38 & 1 & 1.98 & NA \\
    email-enron-only & 143 & 623 & 1.12 & 19 & 31.52 & NA \\
    infect-hyper & 113 & 2196 & 1.22 & 36 & 58.82 & 0.033 \\
    tech-routers-rf & 2113 & 6632 & 5.81 & 1 & 7.41 & NA \\
    road-euroroad & 1174 & 1417 & 6.34 & 19 & 36.74 & NA \\
    \hline
\end{tabular}
\bigskip
\caption{Total solution time in seconds for real world graphs based on a single run. NA indicates that no result was produced after 19 hours of wall clock runtime. \label{table:real_graphs}}
\end{table}

\subsection{Performance on future D-Wave architectures}
\label{sec:future}
\begin{figure}[t]
    \centering
    \includegraphics[width=0.49\textwidth]{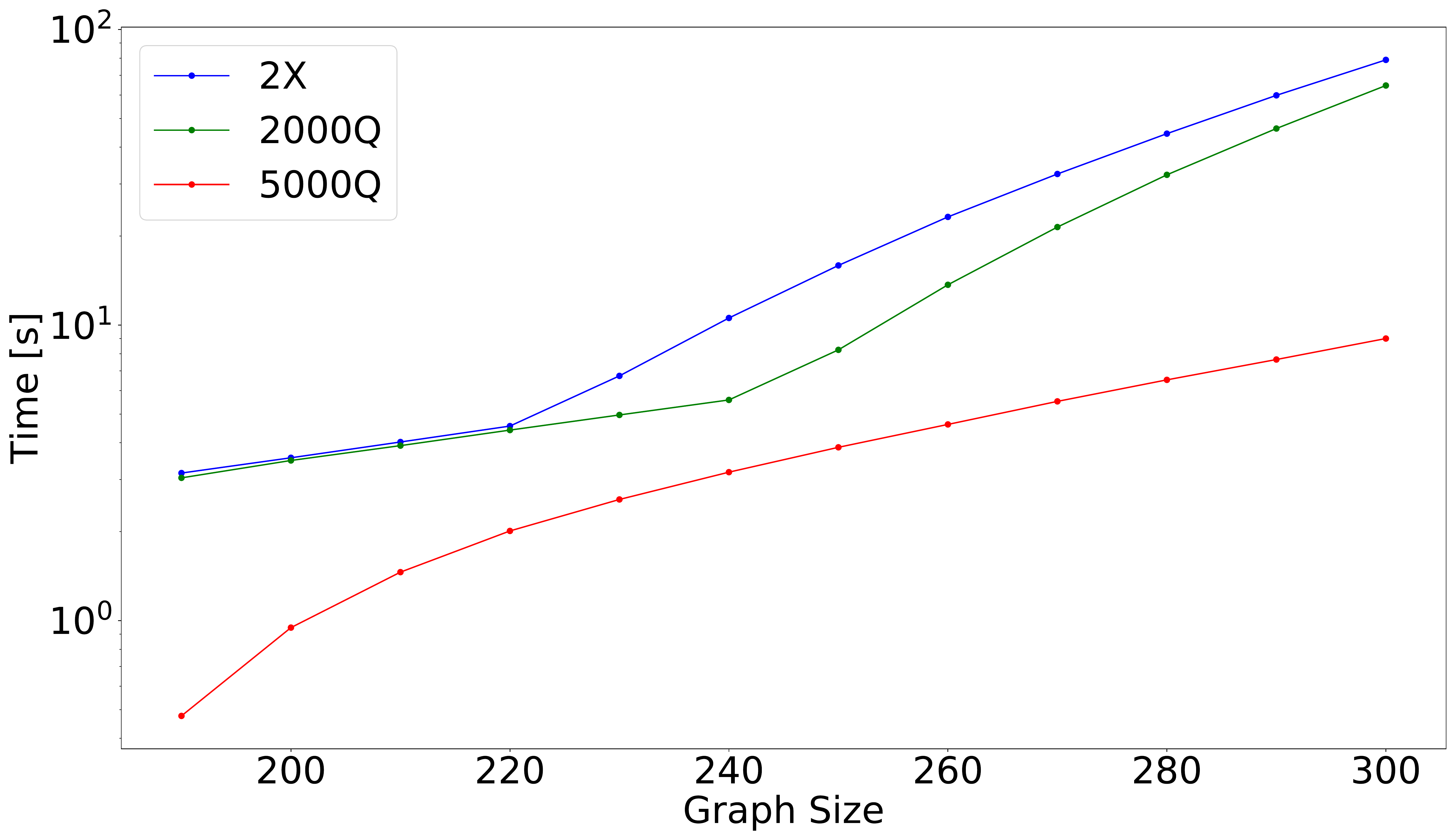}\hfill
    \includegraphics[width=0.49\textwidth]{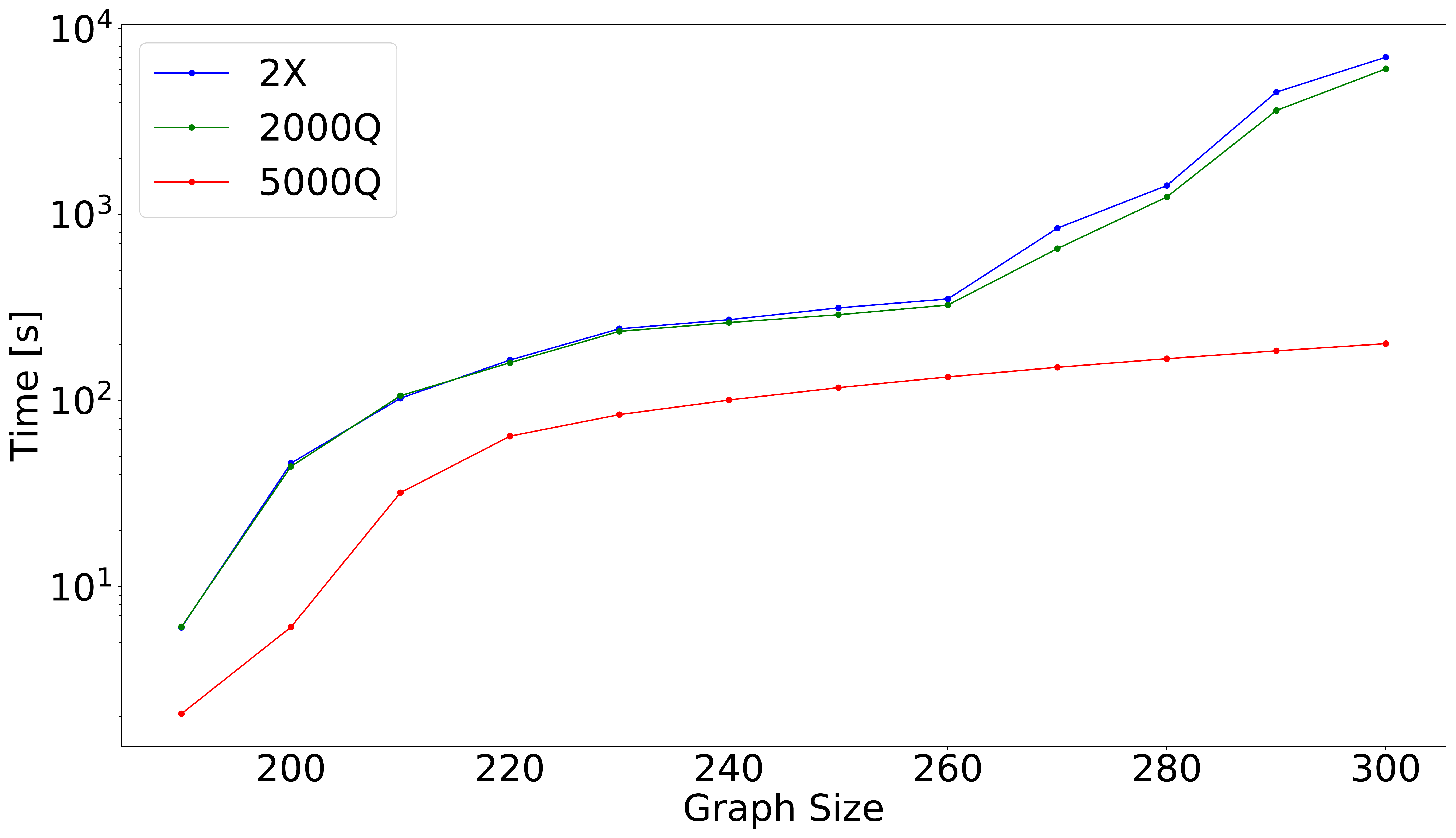}
    \caption{DBR algorithm applied to different D-Wave architectures with fixed average graph degrees of $160$. Preprocessing time (left) and solution time (right). \label{fig:dwave}}
\end{figure}

One way of using the DBR algorithm is to perform the described decomposition until the subgraphs are of a size that can be solved on one of the D-Wave quantum annealers. For D-Wave 2X, the largest graph of arbitrary density that can be embedded onto the quantum processing unit may roughly contain $46$ vertices (this number varies due to manufacturing errors of the chip). For D-Wave 2000Q, arbitrary graphs of roughly $65$ vertices can be embedded. The next generation of D-Wave annealers (currently named \textit{Pegasus P16}) has a higher number of qubits (5640) and a higher connectivity (40484 couplers vs.\ 6000 for 2000Q) than the previous D-Wave machines, resulting in a maximum complete graph size of $180$ that can be embedded onto its quantum processor. We refer to this machine as 5000Q in this paper.

We are interested in how the performance of the DBR algorithm changes for those architectures. Figure~\ref{fig:dwave} shows subgraph count and runtime for the three different architectures (D-Wave 2X, 2000Q, and 5000Q) as a function of the graph size, while we fixed the average degree at $160$ in all generated graphs. All results are averages over $10$ repetitions. The figure shows that D-Wave 2X and 2000Q have a very similar behavior, which is to be expected as both use the same low qubit-connectivity architectures \citep{TechnicalDescriptionDwave}, leading to very similar sizes of the largest embeddable graph on the qubit chip (roughly $46$ vs.\ $65$ vertices) despite the roughly $1000$ and $2000$ nominal qubits, respectively. In contrast, the considerably larger embeddable graph size on the next generation Pegasus architecture leads to a significantly better scaling behavior of our algorithm.

\section{Discussion}
\label{sec:discussion}
This article proposed a novel decomposition algorithm for the MVC problem. The algorithm recursively splits a given MVC instance into smaller subproblems until, at some recursion level, the generated subproblems can be solved with any method of choice, including a quantum annealer. The algorithm is exact, meaning that the optimal solution of MVC is guaranteed given all subproblems can be solved exactly. If subproblems are solved probabilistically (with error probability of at most $\epsilon$), then this guarantee carries over to an error of at most $\epsilon$ for the overall solution returned by our algorithm.

Several pruning and reduction techniques were investigated in order to reduce the number of generated subproblems during the recursion. Our simulation study (a) evaluates possible choices for vertex selection, bounds and reduction techniques for our DBR algorithm and (b) evaluates the DBR algorithm on random and DIMACS graphs as well as on larger D-Wave architectures. We summarize our findings as follows:
\begin{enumerate}
    \item Selecting the highest degree vertex at each recursion level to split each graph further is both fastest in terms of preprocessing time as well as overall solution time. The combination of decomposition upper bound and chromatic number lower bound yields the best trade-off between speed and reduction power. Reduction methods seem to only have an impact at low graph densities which, however, are the most interesting one since they are the hardest cases for our decomposition algorithm, and here the neighbor based vertex removal performs best. We call the algorithm that uses the aforementioned choices in the decomposition approach of Section~\ref{sec:algorithm} the DBR (decomposition, bounds, reduction) algorithm.
    \item The DBR algorithm has good predictable scaling behavior on random graphs of fixed vertex degree. DIMACS benchmark graphs (with as many as thousand vertices and hundreds of thousands of edges), as well as real world graphs from the Network repository (with thousands of vertices and more than ten thousand edges), can be solved exactly with DBR in reasonable time.
    \item One main application of DBR is to split vertex cover instances until they can be solved on D-Wave. The similar architectures of D-Wave 2X and 2000Q exhibit worse scaling behavior than the next generation D-Wave Pegasus P16, which has a much higher qubit connectivity in addition to a higher qubit number.
\end{enumerate}
Future work could include the investigation of further techniques to bound, reduce and prune subproblems as well as an improved implementation of the DBR algorithm. In particular, if it was possible to calculate the Lov\'asz number for large graphs in an efficient manner, using the Lov\'asz lower bound for MVC could greatly improve the capabilities of our decomposition algorithm for solving large MVC problems on D-Wave.


\end{document}